\documentstyle[11pt,newpasp,twoside,epsf]{article} 
\def\edcomment#1{\iffalse\marginpar{\raggedright\sl#1\/}\else\relax\fi} 
\reversemarginpar 

\begin{document} 
\title{Fitting Dynamical Models to Observations of \hfil\break
Globular Clusters}

\author{Dean E. McLaughlin} 
\affil{Space Telescope Science Institute, 3700 San Martin Drive,
Baltimore, MD 21218  USA}

\begin{abstract} 
The basic ingredients of models for the internal dynamics of globular
clusters are reviewed, with an emphasis on the description of equilibrium
configurations. The development of progressive complexity in the models is
traced, concentrating on the inclusion of velocity anisotropy, rotation,
and integrals of motion other than energy. Applications to observations of
extragalactic globulars and to combined radial-velocity and proper-motion
datasets are discussed.
\end{abstract}

\section{Introduction} 

The dynamical modeling of globular clusters has a long and rich history that
intersects repeatedly with fundamental developments in modern
astrophysics. A flavour of this history has already been given by G.~Meylan
in his introduction to this volume, and a fuller account may be found in
the review of Meylan \& Heggie (1997). In this shorter contribution, it is
necessary
to omit discussion of a wide range of interesting and important topics, many
of which are also covered thoroughly by Meylan \& Heggie and updated by other
articles in the present proceedings. Thus, the following does not
consider
any details of the approach of a dense stellar system to a relaxed state,
nor indeed any aspect of the ``microphysics'' of dynamical evolution
(evaporation, core collapse and oscillations, the approach to energy
equipartition, the role of binaries, three-body encounters, and more).
Instead, the focus is on the comparison of dynamical models with only the
most basic of cluster observables: surface-brightness profiles and internal
stellar velocity distributions along the line of sight and in the plane of
the sky. The valuable addition of data on present-day stellar mass functions
is also left to other discussions.
Even in this already limited context, it is necessary to distinguish between
at least three classes of model that are in play.

First is the classic approach pioneered by King (1966a) and Gunn \& Griffin
(1979) (see also Lupton et al.~1987), in which a static cluster is
assumed and a parametric form for the stellar phase-space distribution
function (DF) specified at the outset. Basic equations of stellar dynamics
are then applied to compute self-consistent spatial densities and velocity
moments that can be compared (after projection) to data in order to constrain
the parameters of the assumed DF. Although this approach can be restrictive
in the sense that it forces a functional form on the DF that might not be
appropriate in reality, it is conceptually very well defined, computationally
accessible on any desktop computer, and impressively successful. It has been
applied, at varying levels of complexity, to almost every one of the $\sim$150
globular clusters in the Milky Way, and to many in other galaxies as well.

Second are non-parametric analyses, which still assume a static equilibrium
but begin by constructing smooth and robust functions that describe the
observed run of surface brightness and velocity dispersion. Abel-type
deprojections are then applied to these to derive the intrinsic density and
velocity fields, and an estimate of a phase-space DF whose functional form
is dictated more directly by the data (but not completely so, since
additional, and restrictive, assumptions---such as velocity isotropy---are
always required in the end to extract the DF from deprojected velocities and
space densities). These techniques have been applied to only a few individual
globulars (e.g., Gebhardt \& Fischer 1995; Merritt et al.~1997).

Third, evolutionary models that rely on direct
integrations of the Fokker-Planck equation (a second-order approximation to
the full, time-dependent statistical mechanical master equation for the
stellar DF, which includes the effects of gravitational scattering)
can be used to construct a detailed history of a theoretical cluster, and thus
a sequence of models vs.~time for comparison with observation. This approach
is the most physically rich by far of the three, as it allows potentially for
the inclusion of much of the detailed microphysics mentioned above. It is
computationally intensive, however, and still requires certain assumptions
on the underlying stellar dynamics to be made for practical reasons. It has
only been applied to a handful of clusters (Drukier et al.~1992; Grabhorn et
al.~1992; Drukier 1995; Dull et al.~1997; Giersz \& Heggie 2003).

Aside from these established procedures, it should soon be possible to compare
still more general and realistic types of models with globular cluster
observations; indeed, the amount and variety of surface-brightness,
radial-velocity, proper-motion, and mass-function data now becoming available
for individual clusters fairly demand it. Adaptations of the orbit-based
modeling technique long applied to galaxies (Schwarzschild 1979; Richstone
\& Tremaine 1984) are promising as a new way to static cluster models
that assume nothing {\it a priori} about the form of the stellar distribution
function or the shape of the stellar velocity ellipsoid (e.g., Gebhardt et
al.~2002); and direct $N$-body simulations that require only the input of
correct initial conditions to produce star-by-star models of clusters are
not far off.

The following discussion reviews some aspects of the first class of
models based on parametric prescriptions for the stellar distribution
function.
This might seem almost old-fashioned in the light of ``new horizons'' in
globular cluster astronomy. But insofar as all models should be expected to
fail at some point, progress comes as much from a clear view of their
limitations as from their successes. It is also worth bearing in mind that,
with the HST, the internal structure of globular clusters outside the
Milky Way and beyond the Local
Group can now be observed. The relative simplicity of the current and
foreseeable data on extragalactic clusters makes them similar in many ways
to those that were available for Galactic globulars twenty years ago or
more, and the simple models which provided so much insight then are no less
useful now.

\vfill

\section{Single-mass, isotropic, lowered isothermal spheres} 

All modern parametric models of globular clusters are founded in the
statistical theory of stellar dynamics developed in the first half of the
twentieth century (see, e.g., Spitzer 1987 for a history).  A key concept
in this connection is the relaxation time, $\tau_{\rm relax}$---the
characteristic timescale on which the velocity of a star in a cluster will
change by order of itself as a result of gravitational encounters
(Spitzer \& Hart 1971). The
value of $\tau_{\rm relax}$ is a function of position in a cluster---it
scales fundamentally as $\sigma^3/\rho$, where $\rho$ and $\sigma$ are the
density and rms velocity of the stellar field in the cluster---and is
commonly evaluated both in the core and at the radius containing half
the total cluster light (or mass) in projection.
Figure 1 is a histogram of the core and half-mass relaxation
times in Galactic globular clusters, taken from Harris' (1996) online
catalogue of cluster properties (see also Djorgovski 1993). This graphically
illustrates the well known fact that the inner regions of globular clusters
tend to be dynamically relaxed: $\tau_{\rm relax}$ there is usually shorter
than the age of the systems.

\begin{figure}[!b]
\plotfiddle{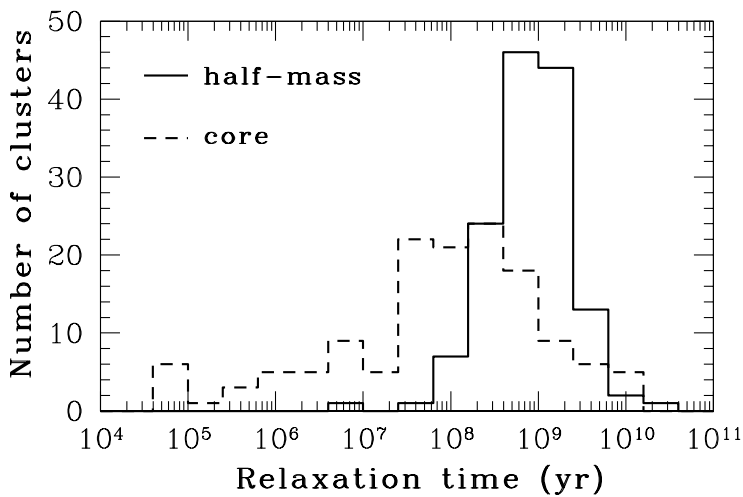}{2.4truein}{0}{60}{60}{-180}{-160}
\caption{Relaxation times of Galactic globular clusters.}
\end{figure}

It is worth recalling, however, that this is {\it not}
generally the case in the outer regions of the clusters (von H\"orner 1957).
If $R_h$ is the
projected half-mass radius of a cluster, then---anticipating the
fact that an average Galactic globular can be described by a King (1966a)
model with a concentration parameter $c\simeq 1.5$---half the cluster mass
is contained in the shell $R_h\la r\la 4R_h$ (still only $\sim 1/3$ of the
way to the tidal radius). The average velocity dispersion
of stars in this shell is perhaps 75\% that of the inner sphere $r\le R_h$,
while the average density of the shell is a factor of 50 lower than that
inside $r\le R_h$. The relaxation time in this outer shell is then $\sim
50\times (0.75)^3 \approx 20$ times longer than that at the half-mass radius;
it becomes comparable to or longer than a Hubble time in most cases.

The importance of relaxation is that it drives the stellar velocity
distribution towards a Maxwellian form, $n(v)\propto \exp(-v^2/2\sigma^2)$,
and on the basis of Fig.~1 this fact must be reflected in any
phase-space distribution function (DF: $f\equiv d^6N/d^3{\mathbf r}\,
d^3{\mathbf v}$) specified as a model for globular clusters.
In addition, Jeans (1915) and Lynden-Bell
(1962a) showed that the DF of a collisionless and steady-state stellar system
(one which is described by the collisionless Boltzmann equation, $df/dt=0$)
is a function only of its isolating integrals of motion (those quantities
that are conserved along a stellar orbit {\it and} that work to confine it to
a finite volume of phase space). There are at most three isolating integrals
for a star
moving in three spatial dimensions: its orbital energy $E$; one derived
from its angular momentum vector $\mathbf{L}$ (e.g., the magnitude $L^2$ in
a spherically symmetric self-gravitating system with an anisotropic velocity
ellipsoid; but only the component
$L_z$ in an axisymmetric and rotating configuration); and a third (say, $I_3$)
which is calculable in closed form for only a handful of simple gravitational
potentials (e.g., Lynden-Bell 1962b; Lynden-Bell 2003).

A third important fact is that a globular cluster moving in the potential of
the Galaxy is necessarily limited in its spatial extent by tidal forces. This
effect is typically included approximately in models by a simple
truncation of the specified DF at high stellar energies corresponding to
motion at the escape velocity  (though see, e.g., Kashlinsky 1988 and Heggie
\& Ramamani 1995 for alternate approaches).

These points were brought together in different ways for models of globular
clusters by many authors, in a flurry of papers in the 1960s (e.g., Oort \&
van Herk 1959; Woolley \& Dickens 1961; Michie 1961, 1963; Michie \&
Bodenheimer 1963; King 1965, 1966a; see King 1966a for a review to
that point).
King's incisive presentation of his particular model in observational terms
was a key factor in its dominance throughout the following decades.

King (1966a) assumed a spherical cluster of identical-mass stars with no
net streaming motions, and an isotropic velocity ellipsoid so that the only
isolating integral of motion is the orbital energy, $E$ [thus, the DF is
$f=f(E)$]. With the further impositions of a near-Maxwellian velocity
distribution for tightly bound stars ($E\ll 0$) and a tidal cut-off
represented by a linear dependence $f\sim -E$ as $E\rightarrow 0$, his
model is defined by $f(E)\propto\rho_0\,\left(2\pi\sigma_0^2\right)^{-3/2}
\left[\exp\left(-E/\sigma_0^2\right) - 1\right]$ for $E=v^2/2-\phi(r) < 0$
(with $\phi=E=0$ defining the tidal radius), and by
$f(E)\equiv 0$ for all $E\ge 0$. The definition
$\rho\equiv\int_0^{\infty} 4\pi v^2 f(E)\,dv$ gives $\rho(\phi)/\rho_0$---the
stellar space density,
normalized by its central value, as a function of the gravitational
potential---and integration of Poisson's equation
$d^2\phi/dr^2+(2/r)d\phi/dr=4\pi G \rho(\phi)$ gives
$\phi/\sigma_0^2$ and then $\rho/\rho_0$ as functions of a dimensionless
radius $r/r_0$. This implicit procedure yields the dimensionless tidal
radius, $r_t/r_0$, at which $\rho=\phi=0$.
It is subject only to the specification of one free boundary
condition: the value $W_0=-\phi(0)/\sigma_0^2$ of the
dimensionless potential at the center of the cluster. Finally, the
one-dimensional velocity-dispersion profile, $\sigma^2/\sigma_0^2$, follows
from $d(\rho\sigma^2)/dr=-\rho\,d\phi/dr$.

A single parameter [$W_0\in(0,\infty)$, or equivalently the concentration
parameter $c\equiv\log(r_t/r_0)$] thus serves to specify fully the
internal structure of a King (1966a) model in dimensionless form. For
comparison with a real cluster, density and velocity profiles as functions
of physical radius are obtained by specifying the additional factors
$\rho_0$, $\sigma_0$, and $r_0$ [only two of which are independent, since
$r_0^2=9\sigma_0^2/(4\pi G \rho_0)$ by dimensional analysis of Poisson's
equation]. Standard
projection integrals applied to these functions yield a directly observable
line-of-sight velocity-dispersion profile, $\sigma_z(R)$ and a surface {\it
mass} density profile $\Sigma(R)$ as functions of projected radius $R$ in
the plane of the sky. Finally, a mass-to-light ratio, $\Upsilon_0$
(spatially constant under the assumption of a single-mass stellar population),
is applied to obtain a model for the observable luminosity surface density
profile: $I(R)\equiv\Sigma(R)/\Upsilon_0$. Thus, the fitting of a King(1966)
model to data ultimately involves four free parameters to be constrained.

Figure 2 shows (following King 1966a) a grid of dimensionless model profiles
in projection. Again, although five different parameters ($c$, and the four
scaling factors) can be counted here, only four are independent given
the necessary connection between $r_0$, $\rho_0$ and $\sigma_0$. Evidently,
fitting a model intensity profile to an observed one is sufficient
to predict the shape of the velocity-dispersion profile, modulo only a
normalization linked to $\Upsilon_0$. Note also that the velocity scale
$\sigma_0$ is not the same, in general, as the central velocity dispersion,
although the two are close in high-concentration (closely isothermal) models.

\begin{figure}[t]
\plotfiddle{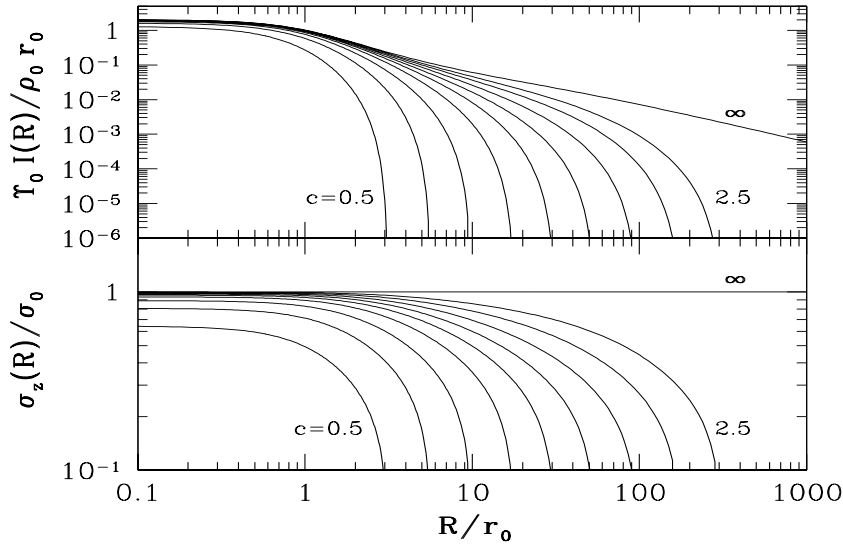}{2.75truein}{0}{60}{50}{-190}{-105}
\caption{Run of luminosity surface density $I$ and line-of-sight velocity
dispersion $\sigma_z$ vs.~projected clustercentric radius $R$ for models
from the family of King (1966a). Profile shapes are set by the
concentration parameter, $c\equiv\log(r_t/r_0)$, with values shown from
$c=0.5$ to $c=2.5$ in steps of 0.25. $c=\infty$ corresponds to a Maxwellian
DF with no high-energy truncation: an
isothermal sphere of infinite spatial extent.}
\end{figure}

It is interesting to consider in slightly more detail the form of the
high-energy cut-off in the otherwise Maxwellian DF specified by King (1966a):
$$f(E)\propto \left\{
\begin{array}{lcl}
\exp\left({-E/\sigma_0^2}\right) - 1 & , & E<0 \cr
0 & , & E\ge 0\ \ ,
\end{array}
\right.$$
which was also adopted by Michie (1963). There is theoretical
justification for this particular cut-off---which leads to $f(E)\sim -E$
as $E\rightarrow 0$---in a relaxed and isolated cluster
whose sphericity is unperturbed by tidal forces (King 1965; Spitzer \&
Shapiro 1971). However (as Fig.~2 illustrates), the exclusion of
high-velocity stars from the DF affects the spatial structure in the outer
halo of a model cluster, a region which in real globulars is neither relaxed
nor unperturbed by tides. Other simple truncations of a single-mass and
isotropic Maxwellian DF have been considered in the literature, such as
(among others) the more sudden one of Woolley \& Dickens (1961):
$$f(E)\propto\left\{
\begin{array}{lcl}
\exp\left(-E/\sigma_0^2\right) & , & E<0 \cr
0 & , & E\ge 0\ \ ;
\end{array}
\right.$$
and the more gradual one of Wilson (1975):
$$f(E)\propto\left\{
\begin{array}{lcl}
\exp\left(-E/\sigma_0^2\right) + E/\sigma_0^2 - 1 & , & E<0 \cr
0 & , & E\ge 0\ \ ,
\end{array}
\right.$$
which gives $f(E)\sim E^2$ as $E\rightarrow 0$.

\begin{figure}[!t]
\plotfiddle{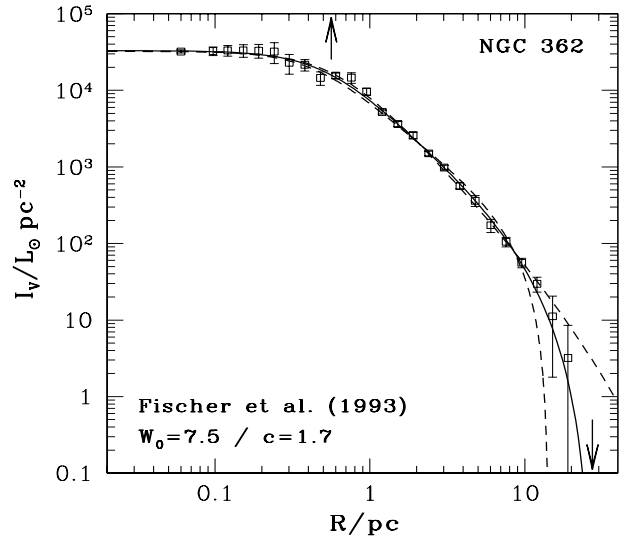}{2.5truein}{0}{50}{45}{-150}{-95}
\caption{Surface brightness profile of NGC 362, as fit by
a King (1966a) model of concentration $c=1.7$ (solid line). Arrows mark
the fitted scale radius $r_0$ and tidal radius $r_t$. Dashed curves are
best fits of self-consistent surface density profiles corresponding to
alternate high-energy truncations of the phase-space distribution function.}
\end{figure}

King (1966a) himself considered each of these alternate energy truncations.
He compared the self-consistent surface-density profiles they predict with
observed star-counts in the halo of M13 and concluded that the data were most
consistent with the linear dependence on $E$ in the high-energy tail of his
DF. Figure 3 illustrates this for a different---and by all appearances
thoroughly average---cluster, NGC 362 (data from Fischer et al.~1993). The
curves, from top to bottom in the right-hand corner of the graph, correspond
to best fits of Wilson's (1975) DF, King's (1966a) DF, and Woolley \& Dickens'
(1961) DF. The standard model is clearly favored, but the
conclusion rests entirely on data from the outermost, least relaxed
and least ``isolated'' parts of the cluster. In cases where a regular
King model might {\it not} provide a perfect description of a
surface-brightness profile (and there are many such cases), it must be
remembered that the shape of the model profile depends not only on the
explicit assumptions of isotropy and equal-mass stars, but also on the
implicit assumptions [through the linear dependence of $f(E)$ near the escape
energy] of relaxation and isolation in the extreme halo.

\begin{figure}[!t]
\plotfiddle{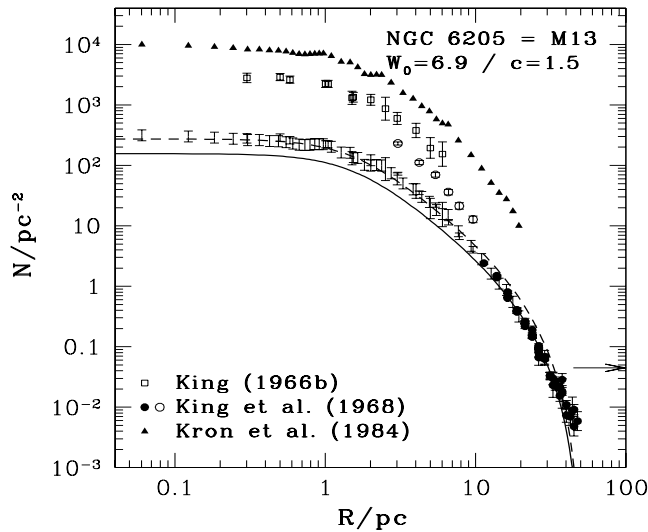}{2.6truein}{0}{50}{45}{-150}{-95}
\caption{Combination of data defining the current empirical surface density
of M13 (NGC\,6205). King (1966a) models plotted both have $c=1.5$, but
different density normalizations and scale radii $r_0$.}
\end{figure}

King and collaborators collected
and fit surface-density data, in the form of star counts and integrated
surface-brightness measurements, for scores of globular clusters (King 1966b;
King et al.~1968). This database was enlarged by other important surveys
over the next twenty years (notably, e.g., Illingworth \& Illingworth 1976;
Kron et al.~1984). All such studies were most recently compiled by Trager et
al.~(1995), who combined often overlapping data to define a
single, calibrated surface-brightness profile and a smooth interpolation
curve for each of 125 Galactic globulars. The example of M13 is illustrated in
Fig.~4. Star counts by King et al.~(1968) in the outer regions of the cluster
define the vertical scale of the graph; another set of inner star counts,
and two sets of photoelectric surface-photometry measurements, are shown with
arbitrary offsets for clarity. To construct a continuous surface-density
profile, the inner star counts and photometry are scaled first to agree with
each other, and then (as indicated by the errorbars with no symbols) to
provide the best match in the (small!) region of overlap with the outer star
counts. Shown as a solid line is the model that King (1966a) found as the
best fit in the outer parts of M13; the dashed curve is the best fit
including the inner data.

Clearly, a King model in this case is not the perfect fit found in NGC 362.
The horizontal arrow at the right of the Figure indicates the level
of background contamination in the star counts of King et al.~(1968), as a
reminder that the observational definition of $I(R)$ is most prone to error
in the outermost regions of the cluster that, as in Fig.~3, are often the
most influential in accepting or rejecting an overall class of dynamical
model. While this caveat must be borne in mind, a more positive aspect of
this plot---which is fairly representative of the profiles collected by
Trager et
al.~(1995)---is the verification that a standard King model is nevertheless
an excellent first-order approximation to the spatial structure of an average
globular cluster. The difference between the solid and dashed curves can be
viewed as an indication of the typical uncertainties in King-model fitting,
which are discussed fully by Trager et al.~(1995). An up-to-date catalogue
of the structural parameters fitting the clusters of Trager et al., and
some 20 others, is maintained online by W.~E.~Harris (Harris 1996).

\begin{figure}[!t]
\plotfiddle{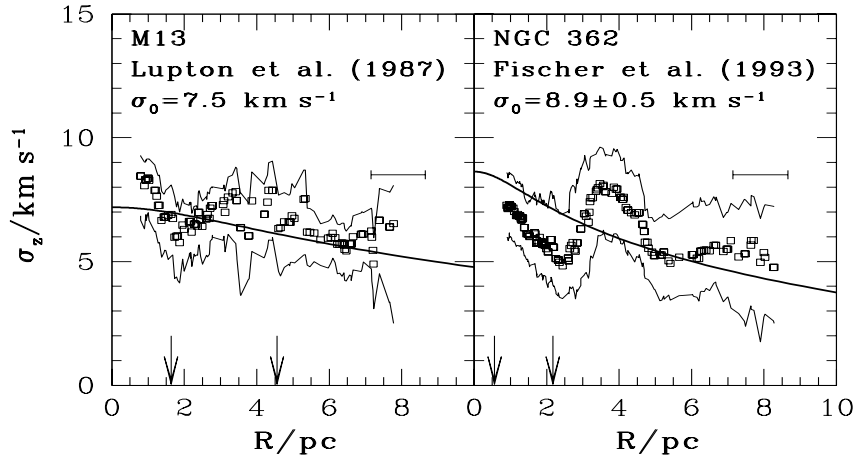}{2.1truein}{0}{60}{50}{-175}{-125}
\caption{Line-of-sight velocity-dispersion profiles
in M13 and NGC 362, compared with single-mass, isotropic, King(1966)
model curves corresponding to the surface-density fits in Figs.~4 and 3.
Arrows mark the fitted scale radius $r_0$ and projected half-light radius
$R_h$.}
\end{figure}

It took somewhat longer for high-quality radial velocities to become available
for comparison with the predicted kinematics of single-mass,
isotropic King models (lower panel of Fig.~2). Illingworth (1976) obtained
integrated-light spectra of the cores of 10 globulars to estimate their
central velocity dispersions and thence (by referring to surface-brightness
fits of a King model) mass-to-light ratios. In terms of direct dynamical
tests, Da Costa et al. (1977) fit models simultaneously to surface brightness
and the velocities of 11 individual giants in NGC 6397, while Gunn \& Griffin
(1979) used radial velocities of 111 giants together with a surface-brightness
profile in M3 to show the need for a multi-mass and anisotropic
generalization of the King model. Their paper opened the floodgates for
such data and analyses (see Meylan \& Heggie 1997 for a complete listing of
the subsequent work), and some 15 years later Pryor \& Meylan (1993) were
able to
summarize velocity information on 56 Galactic globulars. This remains the
most recent such compilation. The sizes of radial-velocity samples in
individual globulars have since grown steadily, reaching now into the
thousands for several clusters.

Figure 5 shows spatially smoothed velocity-dispersion profiles of M13 and NGC
362, constructed from the data published in Lupton et al.~(1987) and
Fischer et al.~(1993), by robustly estimating the rms velocity of all stars
falling within an annulus of fixed width (indicated by the
horizontal errorbars in the Figure) as it is centered on every star in turn.
The jagged lines in the panel for M13 indicate 1-$\sigma$ uncertainties;
those in the NGC 362 panel, 2-$\sigma$ uncertainties. The curves are the
predicted $\sigma_z$ profiles for the King (1966a) models fit to the
surface brightnesses in Figs.~3 and 4, with a normalization $\sigma_0$
corresponding in each case to a $V$-band mass-to-light ratio $\Upsilon_0
\simeq 1.5\ M_\odot\,L_\odot^{-1}$. The basic, {\it first-order} confirmation
of the model, in Figs.~3, 4, and 5 combined, is impressive. However, both
Lupton et al.~(1987) and Gebhardt \& Fischer (1995) conclude through more
detailed analyses of these data---and studies of similar and better datasets
in other clusters also indicate---that non-trivial modifications of King's
single-mass and isotropic assumptions are necessary.

\section{Structural correlations, the fundamental plane, and extragalactic
globulars}

Before outlining the necessary modifications to King (1966a), the general
success of that theory provides an excellent opportunity to construct a
uniform view of the gross structure and internal dynamics of globular
clusters. In particular, the multitude of correlations between fitted
King-model parameters, and any number of physical quantities derived from
them (Djorgovski \& Meylan 1994; Djorgovski 1995; Djorgovski, this volume),
offer deep insight into questions of cluster formation and evolution.

\begin{figure}[!b]
\plotfiddle{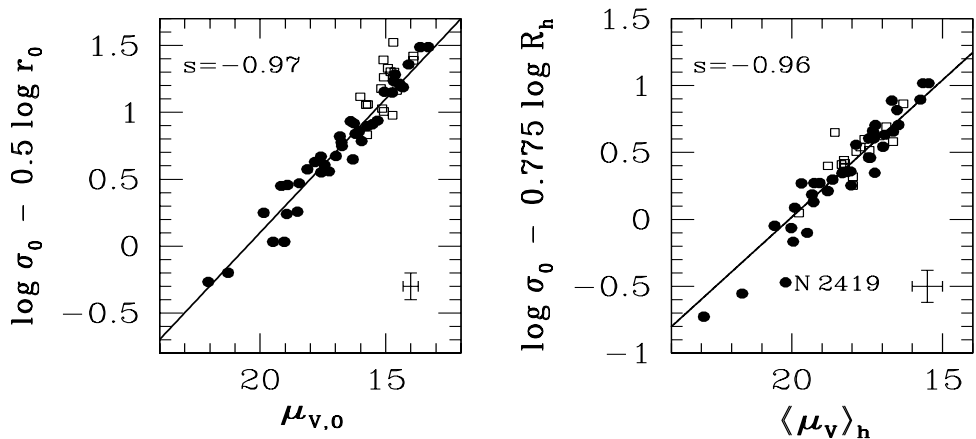}{2.0truein}{0}{70}{60}{-208}{-165}
\caption{One representation of the fundamental plane of Galactic globular
clusters
(cf.~Djorgovski 1995). Solid line in the left panel is the equation of
a uniform mass-to-light ratio $\Upsilon_{V,0}\equiv1.45\ M_\odot\,
L_{V,\odot}^{-1}$ for single-mass, isotropic King-model clusters, as derived
by McLaughlin (2000a). That in the right panel is the equation $E_b=6.6\times
10^{39}\,{\rm erg}\ \left(L/L_\odot\right)^{2.05}$, from McLaughlin (2000a).
Open squares represent core-collapsed clusters.}
\end{figure}

Figure 6 shows the two strongest correlations known for Galactic globulars
with central velocity-dispersion measurements (Djorgovski 1995). The scatter
in these correlations is entirely due to measurement errors, and they define
a fundamental plane analagous to, but significantly different from, that
of elliptical galaxies. The label ``plane'' in this case is entirely accurate
since single-mass, isotropic King models have exactly four free parameters
(\S 2); thus, insofar as real clusters can be considered realizations of
these models, they could in principle occupy the whole of a four-dimensional
space. The existence of two independent correlations restricting their
physical properties instead confines them to a two-dimensional surface inside
this larger space. (In Fig.~6, five different structural parameters are
indicated, but a theoretical King model can always be used to derive any one
of these in terms of the other four.)

Any four quantities can be taken as a physical basis to define the family of
King (1966a) models, so long as they are mutually independent; any two such
sets of variables can be shown to be equivalent, and the choice of
which to work with
is an issue only of convenience or preference. McLaughlin (2000a) takes
the independent cluster properties to be total luminosity $L$, mass-to-light
ratio $\Upsilon_0$, global binding energy $E_b$, and concentration
$c=\log(r_t/r_0)$. He then shows directly that all non--core-collapsed
clusters share a common $\Upsilon_0=1.45\ M_\odot\, L_\odot^{-1}$ in the $V$
band, and obey a strict scaling between total
binding energy, luminosity, and galactocentric position: $E_b=7.2\times
10^{39}\ {\rm erg}\ (L/L_\odot)^{2.05}(r_{\rm gc}/8\ {\rm kpc})^{-0.4}$.
McLaughlin further shows how the correlations in Fig.~6---and all others
between any other cluster properties---derive directly from these
constraints on $\Upsilon_0$ and $E_b$ if all clusters are assumed to
be accurately described by single-mass, isotropic King models. This is the
the physical significance of the fundamental plane.

It is now also possible to fit isotropic King models to extragalactic
globular clusters, and those in M31 (Djorgovski et al.~1997; Barmby et
al.~2002), NGC 5128 (Harris et al.~2002), and M33 (Larsen et al.~2002)
have been found to lie on essentially the same fundamental
plane as those in the Milky Way. The apparently typical scaling between
global energy and luminosity or mass, $E_b\sim L^2 \propto M^2$,
is nontrivial, and understanding it will be important in
developing a comprehensive theory of star and cluster 
formation (McLaughlin 2000b).

\section{Multiple stellar masses, velocity anisotropy, and rotation}

Despite its substantial successes, the simple model of King (1966a) inevitably
showed some cracks when data of very high quality began to accumulate. Figure
7 shows the surface-density profile of M3, as traced by a combination of
star counts and appropriately normalized surface photometry. The two
solid curves are single-mass, isotropic King models of two different
concentrations, which separately describe the inner and outer structure of
this cluster. No single such model is capable of reproducing the full
density profile (as the dashed curve, of concentration intermediate to
the inner and outer fits, suggests).

This point was first made by Da Costa \& Freeman (1976), who suggested that
the problem could be resolved if stars of different masses (say $m_i$ and
$m_j$) had different velocity dispersions at a given radius, in accordance
with energy equipartition ($m_i\sigma_i^2=m_j\sigma_j^2$). They continued to
assume velocity isotropy, and therefore specified a model distribution
function as
$$f(E)=\sum{f_i}\ \ ,\ \ \ \ f_i=C_i\,
\left[\exp\left(-E/\sigma_{i,0}^2\right) - 1\right]\ (E<0)\ .$$
(Oort \& van Herk 1959 also constructed a multimass model for M3, but
with a sharp truncation of the DF at high energies as in Woolley \& Dickens
1961.) The relative numbers of stars in each of several discrete mass
bins---i.e., the values of the coefficients $C_i$---were set by
extrapolation of the observed stellar luminosity funcation,
allowing the construction of a self-consistent model that fit the
surface-density
profile in Fig.~7 well. The allowance of energy equipartition
imparts a larger-than-average velocity dispersion to low-mass stars, thus
providing a greater number of stars on high-energy orbits and an excess
population (relative to the original King model) of the halo regions of the
cluster.

\begin{figure}[!t]
\plotfiddle{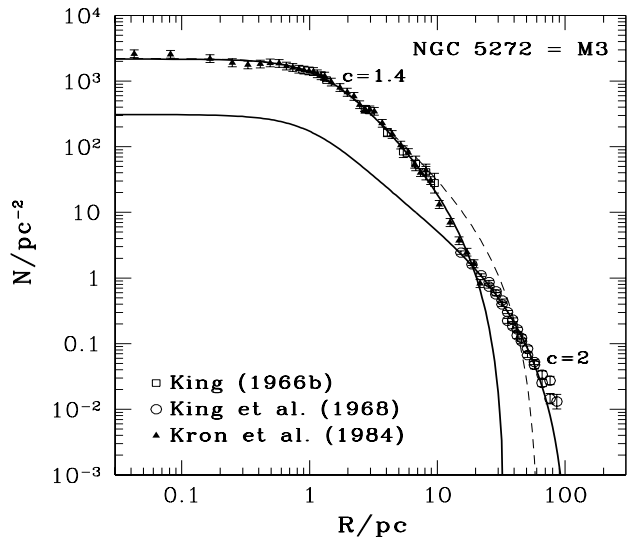}{2.5truein}{0}{50}{45}{-150}{-90}
\caption{Failure of a single King (1966a) model to fit the spatial structure
of M3 (NGC\,5272). After Da Costa \& Freeman (1976).}
\end{figure}

Gunn \& Griffin (1979) revisited M3 with radial velocities for 111 cluster
members also in hand. They argued for a much different extrapolation of the
observed stellar luminosity function and substantially different relative
mass-class populations than Da Costa \& Freeman specified. In this case, they
found it impossible to fit both the surface-brightness profile
and the line-of-sight velocity distribution while keeping the assumption of
isotropy. They continued to assume spherical symmetry (no net rotation) and
defined a ``Michie-King'' model by a DF depending on two isolating
integrals: $E$, and the magnitude $L^2$ of the specific angular-momentum
vector to allow for velocity anisotropy:
$$f(E,L)=\sum{f_i}\ \ ,\ \ \ \ f_i=C_i\,
\exp\left(-L^2/2\sigma_{i,0}^2r_a^2\right)\,
\left[\exp\left(-E/\sigma_{i,0}^2\right) - 1\right]\ (E<0)\ .$$
In the limit $r_a\rightarrow \infty$, this DF reduces to that of Da Costa
\& Freeman; and from that, to the standard King (1966a) form.
More generally, the
functional dependence on $L^2$ in this DF (which was also adopted by Michie
1963 and appears in Eddington's 1915 distribution function) provides for
approximate velocity isotropy inside the sphere $r\la r_a$ and a radial bias
in the velocity ellipsoid that increases toward larger radii. Aside from
this, there is no theoretical motivation for the specific term ${\rm e}^
{-L^2/(2\sigma_{i,0}^2r_a)}$; it is much more {\it ad hoc} than the form
of the high-energy truncation.

Gunn \& Griffin fit all the available data in M3 by a model with $r_a$
comparable to the cluster's projected half-light radius, consistent
with the idea of velocity isotropization by relaxation inside that radius.
The prediction of a radial velocity anisotropy in the outer parts of the
cluster is untested. Meanwhile, it has become routine to fit
surface-brightness and radial-velocity data simultaneously with Michie-King
models, owing in part to
their early application by Pryor and collaborators (e.g., Pryor et al.~1986)
and by Meylan (1987, 1988). Typically, good fits to all available data can
be achieved---hardly surprising given the number of variables in
the models. In fact, the effects of an increase in the relative population
of low-mass stars can mimic those of a more strongly radially biased velocity
ellipsoid, such that several physically different models can often fit
the same dataset equally well.

It is notable that Michie-King fits tend to be
constrained more by features in the surface-brightness profiles (inflection
points, or convexity in the outer regions) than by the radial-velocity
dispersion profiles (which generally do not stray far from their overall
behavior in the isotropic models of Fig.~2).
The obvious danger of inferring velocity anisotropy largely from surface
photometry (or perhaps the inadequacy of the anisotropy term in the
Michie-King distribution function) has been illustrated in $\omega$
Centauri. Merritt et al.~(1997) analyzed, with non-parametric techniques, the
same data that Meylan et al.~(1995) used to fit
Michie-King models to this cluster. Whereas the Meylan et al.~model fits
require strong radial anisotropy setting in at just
under a half-mass radius, Merritt et al.~assumed isotropy throughout the
entire cluster and were able to find a self-consistent DF that
could account for all structural and kinematic data.

The assumption of strict equipartition of energy in these models
may also be questionable, as there are physical scenarios in which multimass
populations of stars in a hydrostatic equilibrium configuration can
{\it not} be in thermal equilibrium (Spitzer 1969; Lightman \& Fall 1978;
Kondrat'ev \& Ozernoy 1982). Michie-King models can provide more
acceptable fits than King (1966a) models given a wider range of more detailed
data in a globular cluster, but at the cost of introducing more free
parameters under more tenuous physical assumptions.

The discovery of significant rotation in some globular clusters (e.g., 47
Tucanae and $\omega$ Centauri: Meylan \& Mayor 1986; M13: Lupton et al.~1987;
and more than 10 others since with $v_{\rm rot}/\sigma$ ranging from
$\sim$0.01 to 0.5) further complicates matters. To take proper
account of rotation, axisymmetry is assumed instead of
spherical symmetry, and the isolating integrals of motion in the stellar
DF are $E$ and the component $L_z$ of angular momentum along the rotation
axis. If arbitrary velocity anisotropy is to be allowed, then the (generally
unknown) third integral must also appear explicitly: $f=f(E,L_z,I_3)$. And
given only radial-velocity data, the inclination of the cluster to the
plane of the sky automatically becomes another free parameter. Despite these
difficulties, Lupton \& Gunn (1987) developed parametric models of multimass,
anisotropic, rotating and flattened clusters by first using $L^2$ as
a (very crude) approximation to $I_3$, and then multiplying every component
of the Michie-King DF by a term of the ({\it ad hoc}) form
${\rm e}^{\gamma L_z}$. Lupton et al.~(1987) were able to fit such
models to the surface brightness and radial velocities in M13, given also
some internal proper motions and the observed stellar mass function
there; but the many enhancements over King's (1966a) model are
again constrained far more strongly by the surface-brightness profile than
by any velocity data (recall Fig.~5 above, which shows the $\sigma_z$ profile
in M13 to be fully consistent with velocity isotropy).

Even with a maximum of approximation and assumption, parametric modeling
at this level of detail is almost prohibitively complex, and the rotating
model of Lupton \& Gunn (1987) has not been applied to any cluster other than
M13. But given the wealth of cluster dynamical
data available now and coming in the near future (large proper-motion
samples; high-quality stellar luminosity functions) it seems inescapable that
accurate two- and three-integral, multimass models will need to be developed.
It may be possible to develop useful parametric, DF-based models by retaining
a reduction to some type of lowered Maxwellian in the appropriate limits
but drawing on the vast literature of galaxy models for revised treatments
of velocity anisotropy and third integrals. However this might
play out, more numerical methods---orbit integrations along the lines of
Schwarzschild (1979), anisotropic Fokker-Planck simulations (Takahashi 1997;
Drukier et al.~1999; Takahashi \& Portegies Zwart 2000), or direct $N$-body
simulations---will necessarily take on new importance.

\section{Proper motions}

The use of internal proper motions in dynamical models has been
understandably more limited than the application of line-of-sight velocities:
a linear velocity dispersion of 10 km s$^{-1}$ or less in a (crowded)
cluster core corresponds, at a heliocentric distance of 8 kpc,
to a dispersion of $\la 0.25$ mas yr$^{-1}$ in proper motion. Large
datasets with time baselines long enough and internal errors small enough to
resolve such motion---let alone trace its variation
with clustercentric radius---are not easily come by from the ground. Notable
efforts by Cudworth and collaborators (e.g., Cudworth 1976a, 1976b, 1979;
Peterson \& Cudworth 1994) allowed for basic velocity-dispersion estimates
and astrometric distance determinations in a handful of clusters. Leonard
\& Merritt (1989) and Leonard et al.~(1992) used limited
proper-motion data (comparable to the earliest radial-velocity data) to
constrain Jeans-equation mass estimates of M35 and M13, while Lupton et
al.~(1987) compared their DF-based model of M13 with the sparse proper
motions available there.

A substantial breakthrough has come recently, both with the publication of
a huge set of ground-based proper motions for nearly 8000 stars in
$\omega$ Centauri by van Leeuwen et al.~(2000), and with the application
of new techniques for space-based astrometry with HST to obtain tens of
thousands of proper motions in $\omega$ Centauri and 47 Tucanae (see King \&
Anderson 2001, 2002; Anderson, these proceedings; and the poster of
McLaughlin et al.~in this volume for reports on preliminary results). The
potential of these data to constrain any kind of detailed dynamical model
is immense, but they have not yet been fully exploited.

Just one taste of the wealth of information contained in
proper motions is provided by Fig.~8, using the data of van Leeuwen et
al.~(2000) for $\omega$ Centauri. Plotted there is the ratio of the
plane-of-sky velocity dispersion in the direction along the projected radial
vector from the center of the cluster ($\sigma_R$) to that in the
projected azimuthal direction ($\sigma_\Theta$). In a spatial average, and
under the assumptions of spherical symmetry and no net rotation (both of
which are, strictly, incorrect here),
this ratio is related to the ratio of
unprojected radial and tangential velocity dispersions,
$\sigma_r/\sigma_\theta$, by (Leonard \& Merritt 1989)
$$2\,\sigma_r^2/\sigma_\theta^2\ =\ 
3\,\sigma_R^2/\sigma_\Theta^2\ -\ 1\ .$$
While this particular relation is not completely general, it is clear that
the usual velocity anisotropy parameter, $\beta\equiv
1-\sigma_\theta^2/\sigma_r^2$, can be deduced directly from proper-motion
dispersions. These data specifically appear to confirm that the strong radial
anisotropy inferred to obtain outside a half-light radius in $\omega$
Centauri (cf.~\S4) according to fits of parametric Michie-King models, may be
more an artifact of the model construction than a physical effect (see also
King \& Anderson 2002).

\begin{figure}[!t]
\plotfiddle{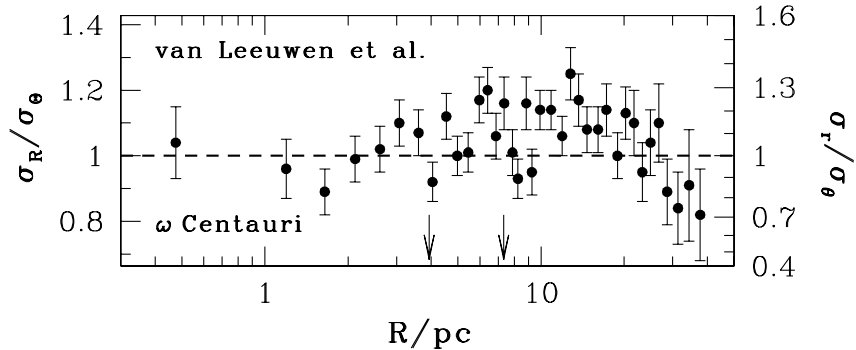}{1.5truein}{0}{60}{60}{-190}{-115}
\caption{Ratio of radial and tangential velocity dispersions on the plane
of the sky in $\omega$ Centauri (NGC\,5139), and the corresponding unprojected
quantity, as a direct measure of velocity anisotropy. The King-model scale
radius and projected half-light radius are indicated.}
\end{figure}

The signature of rotation in proper motions can also be used to estimate
directly the inclination of an axisymmetric cluster, and to delineate
empirically the form of a (non--solid-body) rotation field (e.g., van Leeuwen
et al.~2000). Wybo \& Dejonghe (1995, 1996) and Heacox (1997, 1998) give much
more detailed theoretical discussions on the extraction of dynamical
information from proper-motion datasets. Further development of these and
similar tools, both to mesh them with the new, high-quality data and to extend
them to apply to general gravitational potentials and arbitrary stellar
distribution functions, presents an important challenge.

\end{document}